\documentclass{article}

\normalbaselines
\begin{document}
\begin{center}
\vspace*{1.0cm}

{\LARGE{\bf Globally singularity free semi-classical wave functions in closed
form}}

\vskip 1.5cm

{\large {\bf C. Jung, F. Leyvraz and T.H. Seligman }}

\vskip 0.5 cm

Centro Internacional de Ciencias \\
and\\
Centro de Ciencias F\'\i sicas, University of Mexico (UNAM) \\
Cuernavaca, Mexico

\end{center}

\vspace{1 cm}

\begin{abstract}
We use a factorisation technique and representations of canonical
transformations to construct globally valid closed form
expressions without singularities of semi-classical wave functions
for arbitrary smooth local potentials over a one-dimensional
position space.
\end{abstract}

\vspace{1 cm}

\section{Introduction}
In semi-classical treatments we construct approximate solutions
 of a quantum mechanical problem from the knowledge of the
solutions of the corresponding classical problem.
The usual WKB procedure yields excellent results sufficiently far away from
turning points. Yet at turning points this approximation
leads to singularities. This occurs for all energies below the maximal
value of the potential. Furthermore degenerate situations occur at maxima
of this potential. Some kind of regularisation (uniformisation) is used near
such points. The most common ad hoc solution uses Airy functions near
turning points and Pearcy functions near maxima \cite{co}.

The purpose of this paper is to present a semi-classical global approximation
without singularities. To achieve this
we take advantage of the following facts:

First, the solution of a problem linear in momentum such as
$
p + V(q) - E =0
$
is trivial under canonical quantisation.
Second, the usual Hamiltonian $ H = p^2/2 +V(q)$ can be factorised into two
factors of the first type. Third one of these factors can be converted
to a simple momentum by a canonical transformation.
For quantum mechanics we need Fourier and gauge
transformations \cite{jk}. Their composition introduces errors of order
$\hbar^2$
thus leading to an integral representation of an approximate solution. A
 judicious selection of the integration path depending on the
 coordinate yields converging integrals everywhere, thus
guaranteeing a uniform approximation.

We shall show, that the saddle point approximation of the integral again
yields the WKB solution, and indicate how the path of integration
has to be laid to avoid singularities at turning points and extremal points
of the potential.
We will restrict our discussion to the one dimensional case and a Hamiltonian
with
local  potential, where the action for a given energy $E$
is defined as
\begin{equation}
S(q) = \int^q ds [2(E-V(s))]^{1/2}
\end{equation}

\section{The factorisation}
We start by introducing
the function $\Omega(q,p,E) = H(q,p)-E$. Then the wave function is an
eigenfunction with eigenvalue 0 of the operator version of $\Omega$
under the usual canonical quantisation. If $\Omega$ were linear in $p$,
 we could bring it into the form
\begin{equation}
\Omega = p - g(q,E)
\end{equation}
where $g$ is the derivative of the action function (which would have only
one branch in this case) and the exact wave function
would have the simple form
\begin{equation}
\psi(q)= c \,\exp[i S(q)/ \hbar]
\end{equation}
For Hamiltonian functions of the standard form $\Omega$ is a polynomial
of second degree in p and after multiplication by 2 we can decompose it
into the linear factors
\begin{equation}
\Omega = [p-g(q,E)] * [p+g(q,E)]
\end{equation}
where the two functions $g$ are the derivatives of the two branches of
the action function which differ only in their sign.
It is again trivial to construct eigenfunctions with
 eigenvalue 0
 for each factor. But this does not solve the entire problem because of
the ordering problem of quantum mechanics. In order to obtain an hermitian
operator we must write the factors in some symmetrical order, and there are
infinitely many ways to do so. Fortunately they only
differ in $\hbar^2$ and higher orders in $\hbar$.
Since we want to construct semi-classical
solutions all these possibilities are equally correct for our purpose and
we can select the one which is most convenient. We choose
\begin{equation}
\Omega = [p-g(q,E)]^{1/2}*[p+g(q,E)]*[p-g(q,E)]^{1/2}
\end{equation}
Next we remember, that for canonical transformations which are represented by
gauge
transformations, this representation is quantum mechanically precise \cite{jk}.
At this stage it helps to apply the following canonical transformation
\begin{equation}
p \rightarrow p+g(q,E),\,\, q \rightarrow q
\end{equation}
to bring $\Omega$ into the form
\begin{equation}
\Omega = p^{1/2} [p+2\,g(q,E)] p^{1/2}
\end{equation}
Now assume that $\tilde\chi(p)$ is the Fourier transform of the wave
function $\chi(q)$,
which is eigenfunction with eigenvalue 0 of the operator version of
$[p+2\,g(q,E)]$.
Then
$
\tilde\phi(p) = p^{-1/2} \tilde\chi(p)
$
is eigenfunction with eigenvalue 0 of the operator version of $\Omega$
of Eq.~7 in momentum representation.

\section{Construction of the wave function}
To obtain an expression for the final wave function $\psi(q)$ we have
to assemble all steps in reverse order. First it is obvious that the
function $\chi(q)$  has the form
\begin{equation}
\chi(q) = c \exp[i ( 2\,S(q,E)/ \hbar]
\end{equation}
such that its Fourier transform
\begin{equation}
\tilde\chi(p) = c \int dr \exp[ i (rp+2\,S(r,E)/ \hbar]
\end{equation}
yields $\tilde\chi(p)$ as used above,
Here, and in what follows, we put all uninteresting factors into the
normalisation constant $c$. According to Eq.~8 we obtain
$\tilde\phi(p)$ and next $\phi(q)$ by an inverse Fourier transform as
\begin{equation}
\phi(q)=c \int dp \int dr p^{-1/2} \exp[i (-qp+rp+2\,S(r,E))/ \hbar]
\end{equation}
The p integral can be done in closed form giving
\begin{equation}
\phi(q)=c \int dr \mid q-r \mid^{-1/2} \exp[2i\,S(r,E)/ \hbar]
\end{equation}
Now we make a substitution of the integration variable introducing $s$ as
new integration variable according to
$
r = q - s^2
$
arriving at
\begin{equation}
\phi(q)=c \int ds \exp[2i\,S(q-s^2,E)/ \hbar]
\end{equation}
Finally to arrive at $\psi(q)$ we must undo the canonical transformation
Eq.~6. This is done by multiplying the wave function by the gauge factor
$
\exp[-i\, S(q,E) / \hbar]
$
and we obtain
\begin{equation}
\psi(q)=c \exp[-i\, S(q,E)/ \hbar] \int ds \exp[2i\,S_2(q-s^2,E)/ \hbar]
\end{equation}
This is the formal global solution we wished to obtain, once we determine the
path
of integration. Yet for formal manipulations the fact that we have this closed
form
may be quite important. For example we can readily see that we retrieve the WKB
approximation by a power expansion, as long as we are far away from any turning
point. Then we
formally expand $S(q-s^2,E)$ in a power series in $s^2$ and only keep
the first two terms giving $ S(q,E)- s^2 \partial S(q,E)/ \partial q $.
Plugging in into Eq.~21 gives
\begin{eqnarray}
\psi(q) & = & \exp[-iS(q,E)/\hbar] \int ds \exp[2i s^2 \partial S(q,E)/
\partial q /\hbar] \nonumber\\
& = & c (\partial S(q,E)/ \partial q)^{-1/2} \exp[-iS(q,E)/ \hbar]
\end{eqnarray}
which is the usual WKB solution. To arrive at this result we have taken
the stationary phase contribution of the point $s=0$ to the $s$ integral.
The exponent can have further stationary points at values $s_c$ of $s$
such that $q-s_c^2$ is a
turning point. However, in general, $S$ varies as $(s-s_c)^{3/2}$
in the vicinity of such argument values; therefore its second derivative
goes as $(s-s_c)^{-1/2}$.
Accordingly, the contribution of such points to the saddle
point evaluation of the integral has weight zero. Therefore the point
$s=0$ is the only point giving contributions in saddle point evaluation
of the integral.

\section{The integration path}

We have started from a second order differential equation. Therefore we
must be able to obtain two linearly independent solutions. One way is to
reverse the sign of $S$, the other is by appropriate
choices of the integration path for the variable $s$ in the complex plane.
The integrand is the exponential of some function $f(s)$, the square bracket
in Eq.~20. Along some sectors for the angle $\alpha$ of the complex
variable $s$ the function $f(s)$
acquires large negative real parts and the integrand decays exponentially.
Let us call these intervals $I_j$.
For the sectors in between the integrand explodes exponentially. An
appropriate choice for the integration path is to come in from infinity
in one angle sector $I_{in}$, to pass near the origin and to return to infinity
in a different sector $I_{out}$. Some combinations of the two intervals
will produce the same solution, and some the solution identically
zero. But there should be two different choices leading to two different
solutions. In general the function $f(s)$ can have isolated singularities,
whose position depends on $q$ and $E$. Then we may eventually deform the
integration path into one, encircling some of these singularities.

By a good choice of the integration path we can also be sure that the
solution does not have singularities. To understand this, let us
fix a value of $E$ and consider $q$ in a small neighbourhood of an
arbitrary fixed point $q_0$. Assume for the moment that the potential
$V(q)$ is an analytic function.
Then the integrand may have some singularities
in isolated points in $s$, but outside of them it is analytic.
We call the singular points  $s_j$. As we vary
$q$ the singular points in $s$ will also move in general but remain
in small
neighbourhoods of $s_j(q_0)$. When we choose the integration path such that
it avoids all these little neighbourhoods, then we obtain a function
$\psi(q)$ which is analytic in the neighbourhood of the point $q_0$.
When we vary $q$ over large ranges, then we have eventually to shift the
integration path accordingly to avoid singularities.

The two most important situations where we need the explicit path of the
integral
are near the maximum of potentials and near turning points.
It can be shown that in the first case, if we use the quadratic approximation
for the extremum we retrieve the exact solution, {\it i.e.} the Pearcy
function.
In the second case it does not seem easy to find a path that yields the exact
solution, but by choosing a path that fulfills the above conditions, we obtain
a
solution that has no singularities and numerical inspection shows it
to be quite close to the Airy function.

\section{Conclusions}

We have obtained an integral representation for a semi-classical
approximation, of the wave-function of a standard Hamiltonian with local
potential.
The method involves a factorisation, which causes errors in higher orders of
$\hbar$, which is acceptable for a semi-classical approximation. As we may
expect
the usual WKB method
results from a saddle-point approximation of this integral. Judicious choices
of the path of the integral in the complex plane lead to approximate solutions
which
have no singularities and thus converge everywhere adequately. The method may
be
readily generalized to Hamiltonians which are of higher order in the momenta.
In particular examples it might be useful to modify the method slightly
by multiplying and dividing by additional factors. This can in many cases
provide the exact solution for all values of the energy. In a future
publication we will present such examples in detail.

The advantages of this method consist in the fact that the points where WKB
breaks
down are not treated piecemeal, but are covered by the same integral
representation.
Whenever we wish to make an analytic statement about semi-classics this can
be a great advantage.
\section*{Acknowledgments}
 This work is supported by the UNAM DGAPA project IN-102597.

\end{document}